\documentclass{article} 
\usepackage{iclr2026_conference,times}

\usepackage{amsmath,amsfonts,bm}

\def\eqref#1{equation~\ref{#1}}

\def\1{\bm{1}}

\DeclareMathAlphabet{\mathsfit}{\encodingdefault}{\sfdefault}{m}{sl}
\SetMathAlphabet{\mathsfit}{bold}{\encodingdefault}{\sfdefault}{bx}{n}

\usepackage{hyperref}
\usepackage{url}
\usepackage{graphicx}
\usepackage{subcaption}
\usepackage{booktabs}
\usepackage{algorithm}
\usepackage{algorithmic}

\title{Particle-Guided Diffusion for Gas-Phase\\ Reaction Kinetics}

\author{Andrew Millard  \\
Department of Computer and Information Science\\
Linköping University\\
Linköping, Sweden\\
\texttt{andmi73@liu.se} \\
\And
Henrik Pedersen \\
Department of Physics, Chemistry and Biology \\
Linköping University\\
Linköping, Sweden\\
}

\iclrfinalcopy 
\begin{document}

\maketitle

\begin{abstract}
Physics-guided sampling with diffusion model priors has shown promise for solving partial differential equation (PDE) governed problems, but applications to chemically meaningful reaction-transport systems remain limited. We apply diffusion-based guided sampling to gas-phase chemical reactions by training on solutions of the advection-reaction-diffusion (ARD) equation across varying parameters. The method generates physically consistent concentration fields and accurately predicts outlet concentrations, including at unseen parameter values, demonstrating the potential of diffusion models for inference in reactive transport.
\end{abstract}

\section{Introduction}


Chemical kinetics \cite{atkinson1984kinetics} is fundamental for understanding reaction mechanisms and underpins applications from atmospheric chemistry to catalysis. Given a set of reaction mechanisms, kinetic rate theory expresses the reaction rate of each chemical species as a function of species concentrations and rate constants \citep{atkins2023atkins}. The resulting system of differential equations governs the temporal evolution of species concentrations and is essential for modelling complex reaction networks such as chemical vapor deposition (CVD) \citep{danielsson2020systematic}. \looseness = -1

Classical numerical solvers for partial differential equations (PDEs), such as finite element methods (FEMs) \citep{monk2003finite, kreiss1974finite}, often require careful design of discretisation schemes and solver parameters to achieve high accuracy. When applied to problems involving broad parameter ranges or repeated parameter sweeps, this can result in significant computational overhead and limited flexibility without repeated recalibration.

Recent work has shown that deep generative models, including diffusion models \citep{DBLP:journals/corr/abs-2505-17351, pmlr-v238-kerrigan24a}, can approximate the complex spatiotemporal structure of PDE systems. Building on this foundation, guided sampling methods using diffusion-based generative priors have been proposed to produce physically admissible PDE solutions, with successful applications to systems such as the Navier-Stokes equations and Darcy flow \citep{huang2024diffusionpdegenerativepdesolvingpartial, millard2026particleguideddiffusionmodelspartial}. However, most existing studies focus on idealised benchmark problems, and applications to realistic, parameter-varying physical systems remain limited.

In this work, we demonstrate that diffusion-based guided sampling, when trained on data spanning a range of parameters, can generate accurate solutions to complex PDE systems at previously unseen parameter values. We further show that this approach enables the simulation of gas-phase reaction kinetics across multiple temporal scales and allows for accurate estimation of spatiotemporal concentration fields from sparse observations.

\section{Background} \label{sec:background}

\textbf{Time-Varying Partial Differential Equations.} We consider time-dependent PDEs. For a spatial domain $\Omega$ and a time horizon $[0,\mathcal{T}]$, a PDE can be written as
\begin{equation}
    \begin{split}
        f(u, c, \tau,a) &= 0, \quad c \in \Omega,\quad \tau \in [0,\mathcal{T}], \\
        u(c,\tau) &= g(c,\tau), \quad c \in \partial \Omega,\quad \tau \in [0,\mathcal{T}], \\
        u(c,0) &= a(c),
    \end{split}
\end{equation}
where $c$ denotes a spatial coordinate, $\tau$ denotes physical (PDE) time, $u \in \mathcal{U}$ is the PDE solution field, $a \in \mathcal{A}$ specifies the initial condition (and, more generally, fixed PDE coefficients) and $g$ are the boundary conditions.

Throughout this work, the abstract solution field $u(c,\tau)$ represents a spatiotemporal physical field. In later sections we denote this field as a concentration field $C(c,\tau)$ governed by a reaction–transport PDE.
The notation $C(\cdot,\tau)$ refers to the spatial concentration field over $\Omega$ at a fixed PDE time $\tau$.

\textbf{Sampling for Generative Diffusion Models.} Generative diffusion models \citep{sohl2015deep, song2021scorebasedgenerativemodelingstochastic, ho2020denoising, cao2024survey, karras2022elucidatingdesignspacediffusionbased} provide a probabilistic framework for sampling from complex, high-dimensional data distributions $p_{\mathrm{data}}(x)$. 

The forward process progressively corrupts $p_{\mathrm{data}}(x)$ with Gaussian noise according to a noise schedule $\sigma_t \in [0,\sigma_{\max}]$, $t \in [0,T]$, yielding a reference distribution $p(x_T) \approx \mathcal{N}(0,\sigma_{\max}^2 I)$. Sampling proceeds by drawing $x_T$ from this reference distribution and reversing the noising process to obtain $x_0 \sim p_{\mathrm{data}}(x)$. The reverse-time dynamics are described by the stochastic differential equation
\begin{equation}
    d x_t
    =
    -2 \dot{\sigma}_t \sigma_t \nabla_x \log p_t(x_t,\sigma_t)\, dt
    + \sqrt{2 \dot{\sigma}_t \sigma_t}\, dW_t,
    \label{eq:reverse_sde_noise_form}
\end{equation}
where $W_t$ is a Brownian motion and $\dot{\sigma}_t$ denotes a time derivative. Following \citet{karras2022elucidatingdesignspacediffusionbased}, the score is estimated using a denoising network $\delta_\theta$ as
\begin{equation}
    \nabla_x \log p_t^\theta(x_t,\sigma_t)
    =
    \frac{x_t - \delta_\theta(x_t,\sigma_t)}{\sigma_t^2}.
\end{equation}

\textbf{Guided Sampling.} We consider the setting in which physical parameters are fixed and known, and the diffusion model prior is trained over realisations of the PDE solution field alone. Accordingly, diffusion samples take the form
\begin{equation}
    x \in \mathcal{X} = \mathcal{U},
\end{equation}
where $x$ represents a discretised realisation of the PDE solution field, which in this work corresponds to a concentration field at a single physical time. Conditioning on observations $y$, we aim to sample from the posterior $p_\theta(x \mid y) \propto p(y \mid x)\, p_\theta(x)$ where $p_\theta(x)$ denotes the diffusion model prior. Sampling is performed by simulating the guided reverse-time SDE
\begin{equation}
    \begin{split}
        d x_t =
        -2 \dot{\sigma}_t \sigma_t \nabla_{x_t} \log p_t^\theta(x_t,\sigma_t)\, dt 
        -2 \dot{\sigma}_t \sigma_t \nabla_{x_t} \log p_t^\theta(y \mid x_t,\sigma_t)\, dt
        + \sqrt{2 \dot{\sigma}_t \sigma_t}\, dW_t,
    \end{split}
    \label{eq:guidance_sde_noise}
\end{equation}
where the additional gradient term provides guidance toward consistency with observations and governing equations. We approximate the reverse process using a guided Euler-Maruyama (GEM) update \citep{millard2026particleguideddiffusionmodelspartial} (switching notation from continuous diffusion time $t$ to discrete diffusion sampling indexes $k$)
\begin{align}
    x_{k-1}
    =
    x_k
    +
    (\sigma_{k-1}^2 - \sigma_k^2)
    \frac{x_k - \delta_\theta(x_k,\sigma_k)}{\sigma_k^2}
    +
    (\sigma_{k-1}^2 - \sigma_k^2)
    \nabla_{x_k} \log \tilde{p}_\theta(y \mid x_k)
    +
    \sqrt{\sigma_{k-1}^2 - \sigma_k^2}\, \zeta,
    \label{eq:EM_denoiser_guided_constant}
\end{align}
with $\zeta \sim \mathcal{N}(0,I)$ and $\tilde{p}_\theta$ denotes a likelihood function whose formulation is detailed in later sections. The time interval $\Delta t$ is implicitly accounted for in the noise scheduler. 
\section{Methodology}

\begin{table}[t]
\centering
\tiny
\caption{Outlet concentration summary averaged over common simulation-seed pairs for SMC and ODE methods. We report the mean $\pm$ standard deviation for all metrics: Root Mean Squared Error (RMSE), Mean Absolute Error (MAE) and outlet concentrations of the chemical species at the final timestep (Pred Final). Best values per species are shown in bold (lowest RMSE/MAE and closest final prediction to GT). The RMSE and the MAE are the mean errors across all time steps.}
\resizebox{\textwidth}{!}{%
\begin{tabular}{llcccc}
\toprule
Method & Species & GT Final & Pred Final & RMSE & MAE \\
\midrule
    & NO  & $9.710 \times 10^{-3}$ 
    & $\mathbf{9.879 \times 10^{-3}} \pm 1.35 \times 10^{-2}$ 
    & $\mathbf{5.445 \times 10^{-4}} \pm 8.90 \times 10^{-4}$ 
    & $\mathbf{1.664 \times 10^{-4}} \pm 2.00 \times 10^{-4}$ \\
SMC & O3  & $4.942 \times 10^{-3}$ 
    & $\mathbf{5.093 \times 10^{-3}} \pm 8.68 \times 10^{-3}$ 
    & $\mathbf{5.517 \times 10^{-4}} \pm 9.52 \times 10^{-4}$ 
    & $\mathbf{1.694 \times 10^{-4}} \pm 1.98 \times 10^{-4}$ \\
    & NO2 & $2.464 \times 10^{-2}$ 
    & $\mathbf{2.447 \times 10^{-2}} \pm 1.50 \times 10^{-2}$ 
    & $\mathbf{6.361 \times 10^{-4}} \pm 8.18 \times 10^{-4}$ 
    & $\mathbf{2.322 \times 10^{-4}} \pm 1.92 \times 10^{-4}$ \\
\midrule
    & NO  & $9.710 \times 10^{-3}$ 
    & $7.020 \times 10^{-3} \pm 9.51 \times 10^{-3}$ 
    & $4.430 \times 10^{-3} \pm 2.73 \times 10^{-3}$ 
    & $1.681 \times 10^{-3} \pm 1.19 \times 10^{-3}$ \\
ODE & O3  & $4.942 \times 10^{-3}$ 
    & $4.699 \times 10^{-3} \pm 7.91 \times 10^{-3}$ 
    & $3.468 \times 10^{-3} \pm 2.17 \times 10^{-3}$ 
    & $1.165 \times 10^{-3} \pm 7.94 \times 10^{-4}$ \\
    & NO2 & $2.464 \times 10^{-2}$ 
    & $1.958 \times 10^{-2} \pm 1.13 \times 10^{-2}$ 
    & $3.467 \times 10^{-3} \pm 2.46 \times 10^{-3}$ 
    & $1.559 \times 10^{-3} \pm 1.24 \times 10^{-3}$ \\
\bottomrule
\end{tabular}
}
\label{tab:outlet_summary_smc_vs_ode}
\end{table}

\textbf{Advection--Reaction--Diffusion Model.} We consider transport-reaction systems governed by advection-reaction-diffusion (ARD) equations \citep{RUBIO200890, ChrisCosner2014DiscreteandContinuousDynamicalSystems}. A diffusion model sample
\[
x \equiv C(\cdot,\tau_n)
\]
represents a discretised concentration field at a single physical time $\tau_n \in [0, \mathcal{T}]$, corresponding to the PDE solution field $u(\cdot,\tau_n)$ introduced in Section~\ref{sec:background}. In an axisymmetric cylindrical reactor of length $L$ and radius $A$, the concentration field
\(
C_s(r,z,\tau)
\)
of species $s$ evolves according to
\begin{equation}
\frac{\partial C_s}{\partial \tau}
+ u(r)\,\frac{\partial C_s}{\partial z}
=
D \nabla^2 C_s
+ R_s(C),
\label{eq:ard_general}
\end{equation}
where $r\in[0,A]$, $z\in[0,L]$, $D$ is the molecular diffusivity, and $R_s(C)$ denotes reaction source terms \citep{doi:https://doi.org/10.1002/0471461296.ch3}. The axial velocity field is prescribed as
\[
u(r)=2U_{\mathrm{avg}}\!\left(1-\left(\frac{r}{A}\right)^2\right).
\]

$U_{\mathrm{avg}}$ is the average cross sectional velocity which we assume is the same for all species. In our experiments, physical parameters in the ARD model are treated as known and fixed and inference is performed only over the concentration fields. 


\textbf{Residual-Based Likelihood for Time-Series Reconstruction.} A time series of concentration fields is reconstructed by applying the guided Euler--Maruyama (GEM) proposal within a sequential Monte Carlo (SMC) framework \citep{chopin2020introduction, del2006sequential, zhao2025generativediffusionposteriorsampling} with a diffusion prior \citep{millard2026particleguideddiffusionmodelspartial} independently at each physical time step $\tau_n$. At each $\tau_n$, the diffusion process is initialised from Gaussian noise and a new SMC-guided denoising procedure is run to reconstruct $C(\cdot,\tau_n)$. At each step, candidate reconstructions of $C(\cdot,\tau_n)$ are evaluated using a likelihood that combines agreement with sparse observations and consistency with the governing ARD dynamics.

We can express our likelihood in terms of the residuals of the sparse observations as well as the PDE equation residuals. Therefore we define \emph{PDE residual} likelihood function: 
\begin{equation}
\log p(y \mid x)
\;\propto\;
-\frac{1}{\sigma_{\mathrm{obs}}}
\frac{1}{N_\mathrm{obs}}
\big\|(C_{\mathrm{obs}} - \mathcal{M}\odot x)\big\|_2^2
\;-\;
\frac{1}{\sigma_{\mathrm{pde}}}
\frac{1}{A L}
\big\|\mathcal{R}_n(\hat C; C^{\tau_{n-1}})\big\|_2^2,
\label{eq:likelihood_final}
\end{equation}
where $\mathcal{M}$ is a binary observation mask, $\hat C$ denotes the physical concentration field corresponding to $x$, $C_{\mathrm{obs}}$ are the sparse observations of the ground truth concentration field, and $\sigma_{\mathrm{obs}}$, $\sigma_{\mathrm{pde}}$ are relative weighting factors. $N_\mathrm{obs}$ are the number of sparse observations. Following \citet{wu2023-TDS}, intermediate diffusion likelihoods are approximated by evaluating the likelihood at the denoised reconstruction
$p_t^\theta(y \mid x_t, \sigma_t) \approx p\!\left(y \mid  x_0 = \delta_\theta(x_t,\sigma_t)\right)
=:
\tilde p_\theta(y \mid x_t)$, which coincides with the true likelihood at time $0$, $\Tilde{p}_\theta(y \mid x_0) = p(y \mid x_0)$. Defining the spatial ARD operator
\begin{equation}
\mathcal{F}(C)
=
u(r)\,\frac{\partial C}{\partial z}
-
D\left(
\frac{\partial^2 C}{\partial z^2}
+
\frac{1}{r}\frac{\partial}{\partial r}
\left(
r\,\frac{\partial C}{\partial r}
\right)
\right)
-
R(C),
\end{equation}
where $R(C)$ denotes the nonlinear reaction source term acting component-wise on the species concentration vector $C$, derived from the chemical reaction mechanism and kinetic rate laws.
The PDE residual at discrete physical times $\{\tau_n\}$ is
\begin{equation}
\mathcal{R}_n(C^{\tau_n};C^{\tau_{n-1}})
=
\frac{C^{\tau_n} - C^{\tau_{n-1}}}{\Delta \tau_n}
+
\mathcal{F}(C^{\tau_{n}}),
\qquad
\Delta \tau_n := \tau_n - \tau_{n-1},
\quad n \ge 1.
\end{equation}

$C^{\tau_{n-1}}$ is taken as the previous reconstructed field. At $\tau_0 = 0$, the concentration field is prescribed as $C^{\mathrm{init}} = 0$ as we assume no chemical species present at this time. 

\section{Experimental Setup and Results} \label{sec:experiments}

\begin{figure}[t]
\centering

\begin{subfigure}{0.9\textwidth}
    \centering
    \includegraphics[width=\linewidth]{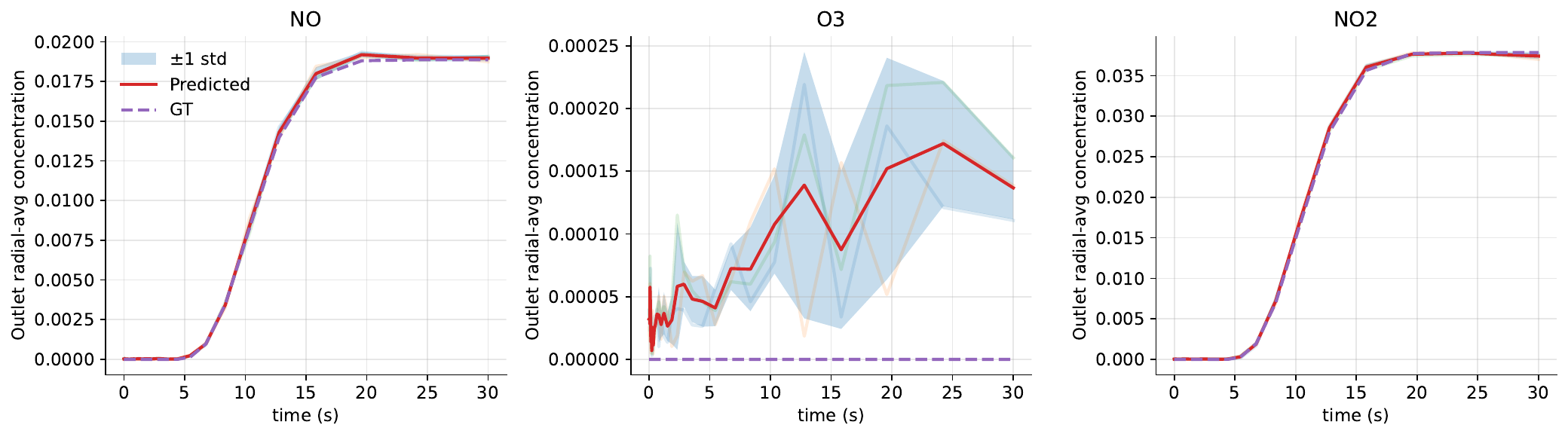}
    \caption{SMC method}
    \label{fig:smc_plot}
\end{subfigure}
\vspace{0.5em}
\begin{subfigure}{0.9\textwidth}
    \centering
    \includegraphics[width=\linewidth]{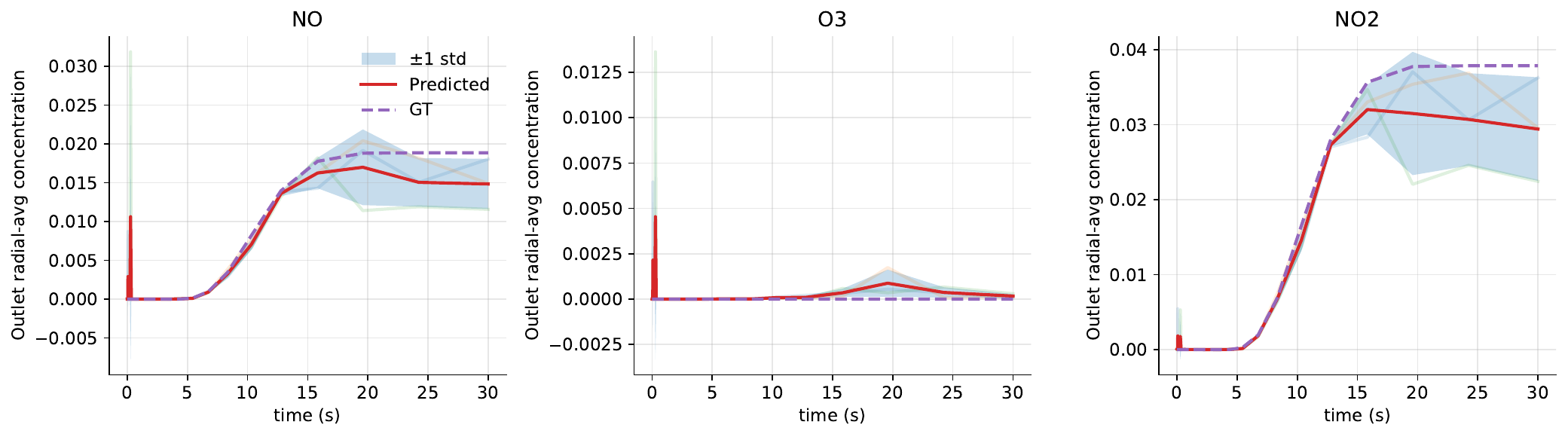}
    \caption{ODE method}
    \label{fig:ode_plot}
\end{subfigure}

\caption{Predicted species concentration at the outlet over time for an example simulation averaged over three seeds. The quantitative results for these simulations are given in Table~\ref{tab:outlet_indiviual_summary}.}
\label{fig:species_outlet}
\end{figure}

\begin{table}[t]
\centering
\tiny
\caption{Quantitative simulation results corresponding to Figures~\ref{fig:species_outlet}. Parameter values are $U_{\mathrm{avg}} = 0.093$, $D = 6.43\times10^{-6} $, $\kappa = 7.51\times10^{3}$. }
\resizebox{\textwidth}{!}{%
\begin{tabular}{llcccc}
\toprule
Method & Species & GT Final & Pred Final & RMSE & MAE \\
\midrule
    & NO 
    & $1.89 \times 10^{-2}$ 
    & $1.90 \times 10^{-2} \pm 1.96 \times 10^{-4}$ 
    & $1.58 \times 10^{-4} \pm 3.38 \times 10^{-5}$ 
    & $8.17 \times 10^{-5} \pm 1.24 \times 10^{-5}$ \\

SMC & O3  
    & $2.24 \times 10^{-9}$ 
    & $1.37 \times 10^{-4} \pm 2.50 \times 10^{-5}$ 
    & $7.41 \times 10^{-5} \pm 9.09 \times 10^{-6}$ 
    & $5.49 \times 10^{-5} \pm 7.75 \times 10^{-6}$ \\

    & NO2 
    & $3.79 \times 10^{-2}$ 
    & $3.74 \times 10^{-2} \pm 3.80 \times 10^{-4}$ 
    & $2.03 \times 10^{-4} \pm 7.11 \times 10^{-5}$ 
    & $1.08 \times 10^{-4} \pm 2.71 \times 10^{-5}$ \\
\midrule
    & NO 
    & $1.89 \times 10^{-2}$ 
    & $1.48 \times 10^{-2} \pm 3.24 \times 10^{-3}$ 
    & $2.90 \times 10^{-3} \pm 2.78 \times 10^{-3}$ 
    & $8.90 \times 10^{-4} \pm 7.88 \times 10^{-4}$ \\

ODE & O3  
    & $2.24 \times 10^{-9}$ 
    & $1.70 \times 10^{-4} \pm 1.18 \times 10^{-4}$ 
    & $1.29 \times 10^{-3} \pm 1.06 \times 10^{-3}$ 
    & $2.73 \times 10^{-4} \pm 2.09 \times 10^{-4}$ \\

    & NO2 
    & $3.79 \times 10^{-2}$ 
    & $2.94 \times 10^{-2} \pm 6.91 \times 10^{-3}$ 
    & $2.81 \times 10^{-3} \pm 1.63 \times 10^{-3}$ 
    & $1.03 \times 10^{-3} \pm 6.25 \times 10^{-4}$ \\
\bottomrule
\end{tabular}
}
\label{tab:outlet_indiviual_summary}
\end{table}

\textbf{Experimental Details.} To evaluate the proposed method, we model the gas-phase reaction
\begin{equation}
\mathrm{NO} + \mathrm{O}_3 \rightarrow \mathrm{NO}_2
\label{eq:no_o3_reaction}
\end{equation}
using a reduced kinetic description involving the species set
\(
s \in \{\mathrm{NO},\,\mathrm{O}_3,\,\mathrm{NO}_2\}.
\)
Assuming elementary mass-action kinetics, the reaction source term
\(
R(C)
\)
appearing in \eqref{eq:ard_general} acts component-wise on the concentration vector
\(
C = (C_{\mathrm{NO}}, C_{\mathrm{O}_3}, C_{\mathrm{NO}_2})
\)
and is given by
\begin{equation}
R(C)
=
\big(
-\kappa C_{\mathrm{NO}} C_{\mathrm{O}_3},\;
-\kappa C_{\mathrm{NO}} C_{\mathrm{O}_3},\;
 \kappa C_{\mathrm{NO}} C_{\mathrm{O}_3}
\big),
\end{equation}
where \(\kappa\) is the second-order reaction rate constant. We consider an axisymmetric cylindrical reactor of radius $a$ and length $L$, with NO and O$_3$ introduced at the inlet and advected downstream at mean velocity $U{\mathrm{avg}}$, where they react to form NO$_2$. Sparse sensors provide partial concentration observations, and the objective is to reconstruct the full concentration fields and estimate outlet concentrations. \looseness = -1

We compare our method to a Guided Euler (GE) discretisation which is used to solve the reverse-time ODE. Details of this can be found in Appendix \ref{app:guided_euler}. Synthetic training and evaluation data are generated by numerically simulating the ARD equations for \eqref{eq:no_o3_reaction} on the cylindrical domain $(r,z)\in[0,A]\times[0,L]$ using a $64\times64$ grid and $32$ logarithmically spaced time snapshots up to $\tau_{\mathcal{T}}=30$ seconds. Physical parameters are randomised across simulations and further details are provided in Appendix~\ref{app:data_generation}. For evaluation, thirty simulations are randomly selected from the test set and the concentration fields are sequentially reconstructed at each physical time step $\tau_n$, with outlet concentrations estimated at each step. GEM is used within an SMC framework \citep{millard2026particleguideddiffusionmodelspartial}. Results are averaged over $3$ independent runs per simulation. A schematic of the experimental setup is also shown in Appendix~\ref{app:data_generation} with further model training and sampling details given in Appendix~\ref{app:model_training_sampling}.  

\textbf{Discussion.} Performance metrics are reported in Table~\ref{tab:outlet_summary_smc_vs_ode}, and predicted outlet concentration time series for an example simulation trajectory are shown in Figure~\ref{fig:species_outlet}. The methods accurately estimate outlet concentrations for the higher-abundance species, while the lower-abundance species exhibit larger relative errors despite maintaining low RMSE and MAE values. Because the governing ARD equations conserve mass, small absolute errors in the higher-concentration species must be compensated by the remaining species, which can lead to large relative errors for species present at very low concentrations when compared to their ground-truth values. We notice that the GEM proposal outperforms the GE method, with the RMSE and MAE errors being an order of magnitude smaller for the former. Appendix~\ref{app:ablation_study} gives an ablation study of the RMSE and MAE with differing number of observations inside the simulated reactor. 

Appendix~\ref{app:time_series} presents reconstructed concentration fields across physical time for the SMC-GEM method for the an example trajectory (corresponding to Figure~\ref{fig:species_outlet} and Table~\ref{tab:outlet_indiviual_summary}) as well as the reactor-wide species fraction change over the simulated period (Figure~\ref{fig:reactor_fraction}). These reconstructions closely match the ground-truth fields, indicating that the larger relative errors observed for O$_3$ are unlikely to be practically significant given the small absolute concentrations. The reconstructed fields further demonstrate the model’s ability to capture gas-phase reaction kinetics across both short and long time scales. Notably, although most reactive interactions occur within the first few seconds before the system approaches equilibrium, the model accurately reproduces the full temporal dynamics even when the data are sampled more densely at early times than at later times. These results also show that our method was able to generalise across a range of different potential parameter values, as none of the exact parameter combinations had been observed in the training dataset. 

These findings demonstrate that guided sampling with diffusion priors can recover physically consistent species concentration fields in reaction-transport systems modelling gas-phase kinetics. The stochastic method produces superior results to the Euler method. The approaches generalise to previously unseen parameter regimes and accurately captures dynamics over extended time horizons.

\section*{Acknowledgment}
This work was partially supported by the Swedish Research Council and the Wallenberg AI, Autonomous Systems and Software Program (WASP) funded by the Knut and Alice Wallenberg (KAW) Foundation. Computations were enabled by the supercomputing resource Berzelius provided by National Supercomputer Centre at Link\"{o}ping University and the KAW foundation.

\section*{Ethics Statement}
This paper presents work whose goal is to advance the field of machine learning through improved methods for scientific modelling of gas-phase reaction kinetics. While machine learning research can have broad societal implications, we do not foresee any direct negative societal impacts arising from this work.

\bibliography{iclr2026_conference}
\bibliographystyle{iclr2026_conference}

\newpage
\appendix
\section{Reproducibility}

\subsection{Guided Euler for Solving Reverse ODEs} \label{app:guided_euler}
Equation~\ref{eq:EM_denoiser_guided_constant} gives the Euler-Maruyama discretisation which is commonly used to solve for the reverse-time SDE. In order to solve the reverse-time ODE: 
\begin{equation}
    \begin{split}
        d x_t =
        - \dot{\sigma}_t \sigma_t (\nabla_{x_t} \log p_t^\theta(x_t,\sigma_t)\ + \log p_t^\theta(y \mid x_t,\sigma_t)) dt,
    \end{split}
    \label{eq:guidance_ode}
\end{equation}
we can use an Euler discretisation which is given by the following: 
\begin{align}
    x_{k-1}
    =
    x_k
    +
    \frac{1}{2}(\sigma_{k-1}^2 - \sigma_k^2)
    \frac{x_k - \delta_\theta(x_k,\sigma_k)}{\sigma_k^2}
    +
    \frac{1}{2}(\sigma_{k-1}^2 - \sigma_k^2)
    \nabla_{x_k} \log \tilde{p}_\theta(y \mid x_k). 
    \label{eq:guided_euler}
\end{align}
Which we refer to as the \emph{Guided Euler} (GE) discretisation method. Further details of the unconditional version of this can be found in \citet{song2021scorebasedgenerativemodelingstochastic} and \citet{karras2022elucidatingdesignspacediffusionbased}. 

\subsection{Pseudocode}

\begin{algorithm}[h!]
\caption{GEM Algorithm}
\label{alg:gem_proposal}
\begin{algorithmic}[1]
\REQUIRE{$\delta_{\theta}(x; \sigma),\, \sigma_k, \sigma_{k-1}, x_{k}, \alpha, y$}
    \STATE Sample $\epsilon_{k} \sim \mathcal{N}(0, I)$
    \STATE $\Theta_{k} \leftarrow \dfrac{x_{k} - \delta_{\theta}(x_{k}, \sigma_{k})}{\sigma_{k}^2}$
    \STATE $x_{k-1} \leftarrow x_{k} + (\sigma_{k-1}^2 - \sigma_{k}^2) (\Theta_{k} + \nabla_{x_{k}}\log \Tilde{p}_\theta(y \mid x_k))  + \sqrt{\sigma_{k-1}^2 - \sigma_{k}^2} \epsilon_{k}$
    \STATE \textbf{return} $x_{k-1}$
\end{algorithmic}
\end{algorithm}

\begin{algorithm}[h!]
\caption{GE Algorithm}
\label{alg:ge_proposal}
\begin{algorithmic}[1]
\REQUIRE{$\delta_{\theta}(x; \sigma),\, \sigma_k, \sigma_{k-1}, x_{k}, \alpha, y$}
    \STATE $\Theta_{k} \leftarrow \dfrac{x_{k} - \delta_{\theta}(x_{k}, \sigma_{k})}{\sigma_{k}^2}$
    \STATE $x_{k-1} \leftarrow x_{k} + \frac{1}{2}(\sigma_{k-1}^2 - \sigma_{k}^2) (\Theta_{k} + \nabla_{x_{k}}\log \Tilde{p}_\theta(y \mid x_k))$
    \STATE \textbf{return} $x_{k-1}$
\end{algorithmic}
\end{algorithm}

\subsection{Synthetic Data Generation Details}

\begin{figure}[!t]
    \centering
    \includegraphics[width=0.5\linewidth]{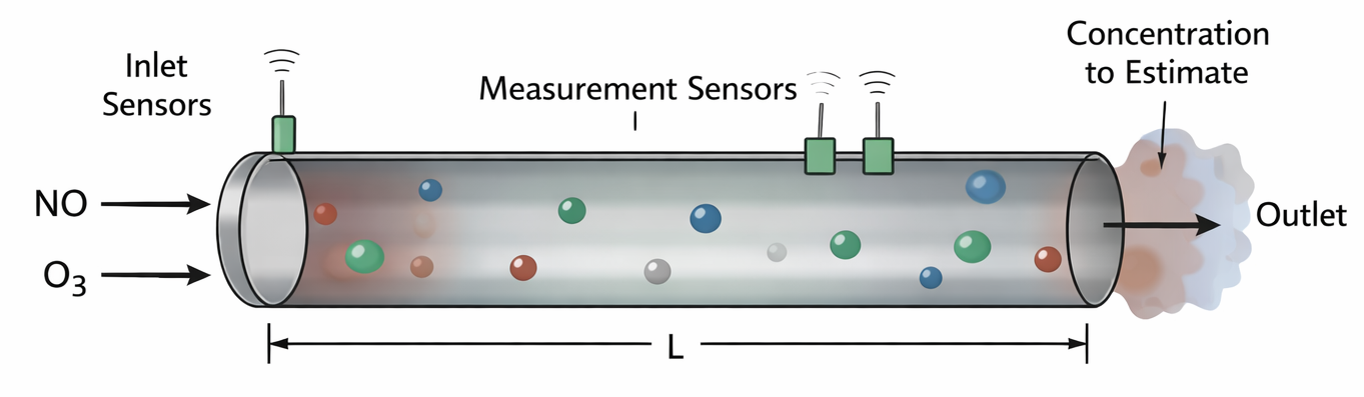}
    \caption{Diagram of the experiment.}
    \label{fig:experiment_diagram}
\end{figure}

\label{app:data_generation}

All synthetic data are generated by numerically solving the transient ARD equation in an axisymmetric cylindrical reactor of length \(L=1.0\) and radius \(A=0.01\). 
The spatial domain \((r,z)\in[0,A]\times[0,L]\) is discretised on a uniform grid with \(N_r=N_z=64\). 
The axial velocity field is prescribed as a Poiseuille profile
\(
u(r)=2U_{\mathrm{avg}}\left(1-(r/A)^2\right).
\)

For each simulation, the mean velocity \(U_{\mathrm{avg}}\) is sampled uniformly from \([0.01,0.5]\) m/s, the molecular diffusivity \(D\) is sampled log-uniformly via \(\log_{10}D\sim\mathcal{U}(-6.5,-4.5)\), and inlet concentrations are sampled as
\(C_{\mathrm{NO,in}},C_{\mathrm{O_3,in}}\sim\mathcal{U}(0,0.05)\) and
\(C_{\mathrm{NO_2,in}}\sim\mathcal{U}(0,0.02)\) mol/m\(^3\).
The reaction rate constant \(\kappa\) is varied by sampling temperature \(T\sim\mathcal{U}(270,320)\) K and computing \(\kappa=\kappa(T)\) using a standard Arrhenius-type expression.

Time integration is performed using an operator-splitting scheme consisting of an explicit upwind advection step in the axial direction, second-order diffusion in the axial and radial directions, and a pointwise update for the bimolecular reaction kinetics. 
Symmetry conditions are enforced at \(r=0\), no-flux boundary conditions at \(r=A\), and Dirichlet conditions at the inlet \(z=0\).
Each trajectory is simulated up to \(\tau_{\mathcal{T}}=30\) seconds, with \(32\) logarithmically spaced snapshots recorded between \(\tau_{\min}=5\times10^{-2}\,\mathrm{s}\) and \(\tau_{\mathcal{T}}\), including the initial condition at \(\tau_0=0\).

Using this procedure, we generate \(2000\) independent simulations, yielding \(64{,}000\) three-channel concentration fields of shape \((3,64,64)\). 
Of these, \(62{,}000\) samples are used for training the diffusion model (a U-Net architecture \citep{ronneberger2015u}) and the remainder were used for evaluation.

\subsection{Model Training and Sampling Details} \label{app:model_training_sampling}
As the diffusion prior, we use build upon the \citet{huang2024diffusionpdegenerativepdesolvingpartial} repository which itself builds upon the \citet{karras2022elucidatingdesignspacediffusionbased} repository. A batch size of 64 was used when training the diffusion prior and a total of 0.5 million gradient evaluations were taken over the dataset during training. 

When sampling, for each method we use $K=400$ discretised steps with the SMC method using 8 samples and the ODE only using 1 as the only randomness is the initial noise the samples are created from and no criteria to choose the individual best sample. $\sigma_{\mathrm{obs}} = 5$ and $\sigma_{\mathrm{pde}} = 1$ were used for the variance parameters for all simulations. 

The average run time for simulating all the trajectories was $2.659\times10^3 \pm 3.89\text{s} \approx 45 \text{mins}$ for SMC method and $1.686\times10^3 \pm 6.17\text{s} \approx 28 \text{mins}$ for the ODE method. We note that the SMC method does incur an extra inference time cost but produces superior results to the ODE method. 

\section{Ablation Results} \label{app:ablation_study}

Table~\ref{tab:outlet_ablation_summary} gives the results of an ablation study for differing numbers of observations inside the reactor cylinder. As expected generally the error decreases as we increase the number of observations from the sensors however, we still report good RMSE and MAE values for lower numbers of observations ($N_{\text{obs}} = 25$ $\approx$ $0.6\%$ of pixels in the data sample).
\begin{table}[H]
\centering
\tiny
\caption{Ablation simulation results for differing numbers of observations inside the reactor when using the GEM proposal with an SMC framework. Mean $\pm$ standard deviation are reported across metrics.}
\resizebox{\textwidth}{!}{%
\begin{tabular}{llcccc}
\toprule
$N_{\text{obs}}$ & Species & GT Final & Pred Final & RMSE & MAE \\
\midrule
      & NO & $1.89 \times 10^{-2}$ & $1.90 \times 10^{-2}$ $\pm$ $2.48 \times 10^{-5}$ & $1.95 \times 10^{-4}$ $\pm$ $6.80 \times 10^{-5}$ & $1.02 \times 10^{-4}$ $\pm$ $2.24 \times 10^{-5}$ \\

25  & O3 & $2.24 \times 10^{-9}$ & $\boldsymbol{9.89 \times 10^{-5}}$ $\pm$ $8.82 \times 10^{-5}$ & $\boldsymbol{4.80 \times 10^{-5}}$ $\pm$ $1.22 \times 10^{-5}$ & $\boldsymbol{3.16 \times 10^{-5}}$ $\pm$ $6.89 \times 10^{-6}$ \\

      & NO2 & $3.79 \times 10^{-2}$ & $3.76 \times 10^{-2}$ $\pm$ $3.83 \times 10^{-4}$ & $4.55 \times 10^{-4}$ $\pm$ $1.55 \times 10^{-4}$ & $2.14 \times 10^{-4}$ $\pm$ $5.30 \times 10^{-5}$ \\

\midrule
      & NO & $1.89 \times 10^{-2}$ & $1.89 \times 10^{-2}$ $\pm$ $8.47 \times 10^{-5}$ & $1.81 \times 10^{-4}$ $\pm$ $3.25 \times 10^{-5}$ & $9.24 \times 10^{-5}$ $\pm$ $1.56 \times 10^{-5}$ \\

50  & O3 & $2.24 \times 10^{-9}$ & $1.24 \times 10^{-4}$ $\pm$ $2.11 \times 10^{-5}$ & $5.53 \times 10^{-5}$ $\pm$ $5.30 \times 10^{-6}$ & $4.05 \times 10^{-5}$ $\pm$ $4.34 \times 10^{-6}$ \\

      & NO2 & $3.79 \times 10^{-2}$ & $3.77 \times 10^{-2}$ $\pm$ $9.47 \times 10^{-5}$ & $3.46 \times 10^{-4}$ $\pm$ $5.85 \times 10^{-5}$ & $1.65 \times 10^{-4}$ $\pm$ $1.95 \times 10^{-5}$ \\

\midrule
      & NO & $1.89 \times 10^{-2}$ & $1.90 \times 10^{-2}$ $\pm$ $3.06 \times 10^{-4}$ & $1.67 \times 10^{-4}$ $\pm$ $2.61 \times 10^{-5}$ & $8.96 \times 10^{-5}$ $\pm$ $1.23 \times 10^{-5}$ \\

100  & O3 & $2.24 \times 10^{-9}$ & $1.13 \times 10^{-4}$ $\pm$ $6.00 \times 10^{-5}$ & $7.19 \times 10^{-5}$ $\pm$ $1.33 \times 10^{-5}$ & $5.50 \times 10^{-5}$ $\pm$ $7.90 \times 10^{-6}$ \\

      & NO2 & $3.79 \times 10^{-2}$ & $\boldsymbol{3.78 \times 10^{-2}}$ $\pm$ $4.39 \times 10^{-4}$ & $2.62 \times 10^{-4}$ $\pm$ $7.94 \times 10^{-5}$ & $1.14 \times 10^{-4}$ $\pm$ $2.90 \times 10^{-5}$ \\

\midrule
      & NO & $1.89 \times 10^{-2}$ & $1.89 \times 10^{-2}$ $\pm$ $8.05 \times 10^{-6}$ & $7.91 \times 10^{-5}$ $\pm$ $1.39 \times 10^{-5}$ & $5.75 \times 10^{-5}$ $\pm$ $1.04 \times 10^{-5}$ \\

200  & O3 & $2.24 \times 10^{-9}$ & $1.11 \times 10^{-4}$ $\pm$ $6.94 \times 10^{-5}$ & $7.43 \times 10^{-5}$ $\pm$ $1.07 \times 10^{-5}$ & $5.85 \times 10^{-5}$ $\pm$ $2.89 \times 10^{-6}$ \\

      & NO2 & $3.79 \times 10^{-2}$ & $3.78 \times 10^{-2}$ $\pm$ $6.98 \times 10^{-5}$ & $1.00 \times 10^{-4}$ $\pm$ $8.26 \times 10^{-6}$ & $6.41 \times 10^{-5}$ $\pm$ $2.63 \times 10^{-6}$ \\

\midrule
      & NO & $1.89 \times 10^{-2}$ & $\boldsymbol{1.89 \times 10^{-2}}$ $\pm$ $1.43 \times 10^{-4}$ & $\boldsymbol{5.85 \times 10^{-5}}$ $\pm$ $8.63 \times 10^{-6}$ & $\boldsymbol{4.57 \times 10^{-5}}$ $\pm$ $5.92 \times 10^{-6}$ \\

300  & O3 & $2.24 \times 10^{-9}$ & $1.43 \times 10^{-4}$ $\pm$ $4.47 \times 10^{-5}$ & $7.48 \times 10^{-5}$ $\pm$ $4.68 \times 10^{-6}$ & $5.98 \times 10^{-5}$ $\pm$ $1.91 \times 10^{-6}$ \\

      & NO2 & $3.79 \times 10^{-2}$ & $3.77 \times 10^{-2}$ $\pm$ $1.23 \times 10^{-4}$ & $\boldsymbol{6.84 \times 10^{-5}}$ $\pm$ $1.21 \times 10^{-5}$ & $\boldsymbol{5.20 \times 10^{-5}}$ $\pm$ $5.67 \times 10^{-6}$ \\

\bottomrule
\end{tabular}
}
\label{tab:outlet_ablation_summary}
\end{table}

\newpage
\section{Time Series Graphs} \label{app:time_series}

\begin{figure}[H]
\centering

\begin{subfigure}{0.48\textwidth}
    \centering
    \includegraphics[width=\linewidth]{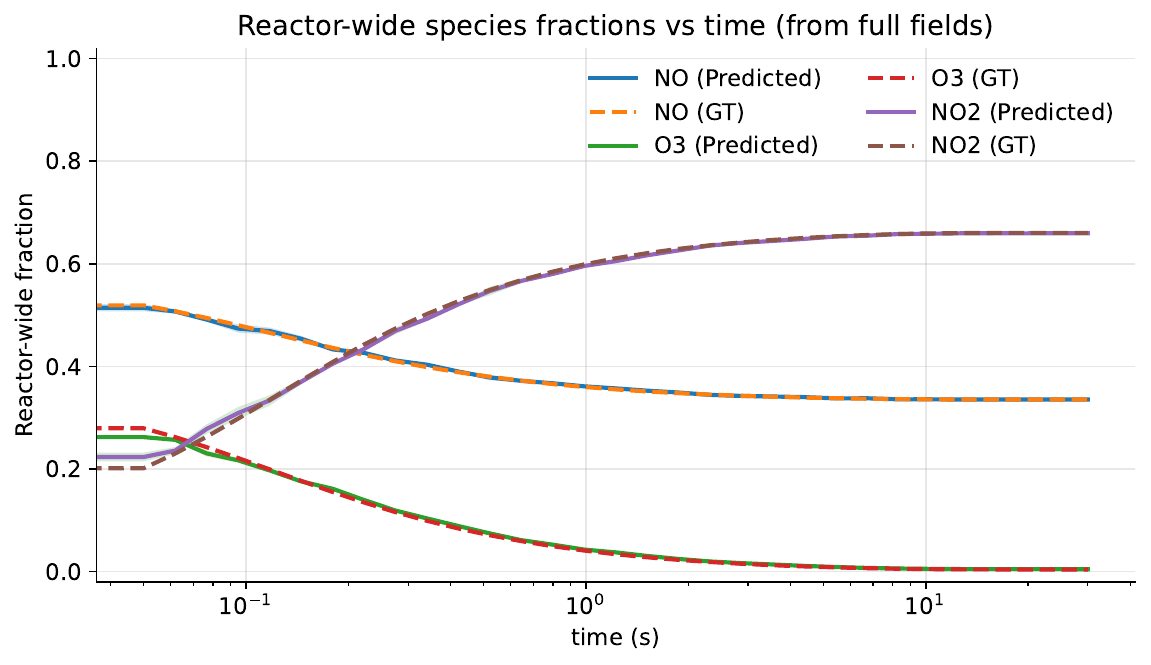}
    \caption{SMC method}
    \label{fig:smc_plot_frac}
\end{subfigure}
\hfill
\begin{subfigure}{0.48\textwidth}
    \centering
    \includegraphics[width=\linewidth]{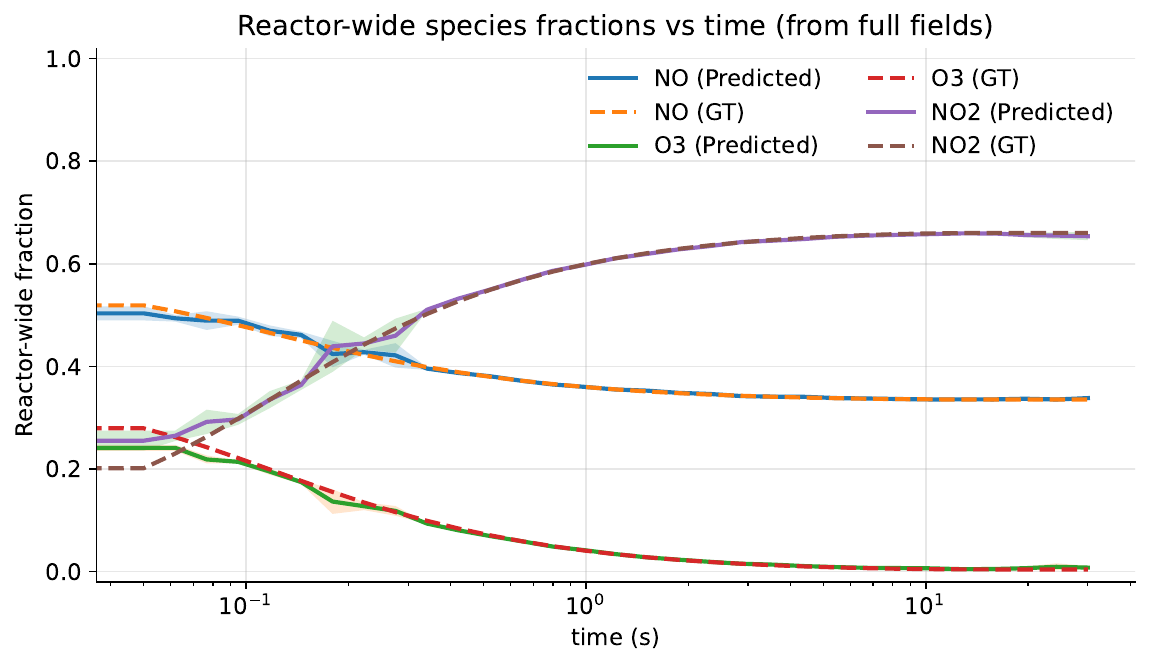}
    \caption{ODE method}
    \label{fig:ode_plot_frac}
\end{subfigure}

\caption{Change in the fraction of species present in the reactor over time.}
\label{fig:reactor_fraction}
\end{figure}

\begin{figure}[H]
\centering

\begin{subfigure}{\textwidth}
  \centering
  \includegraphics[width=0.8\linewidth]{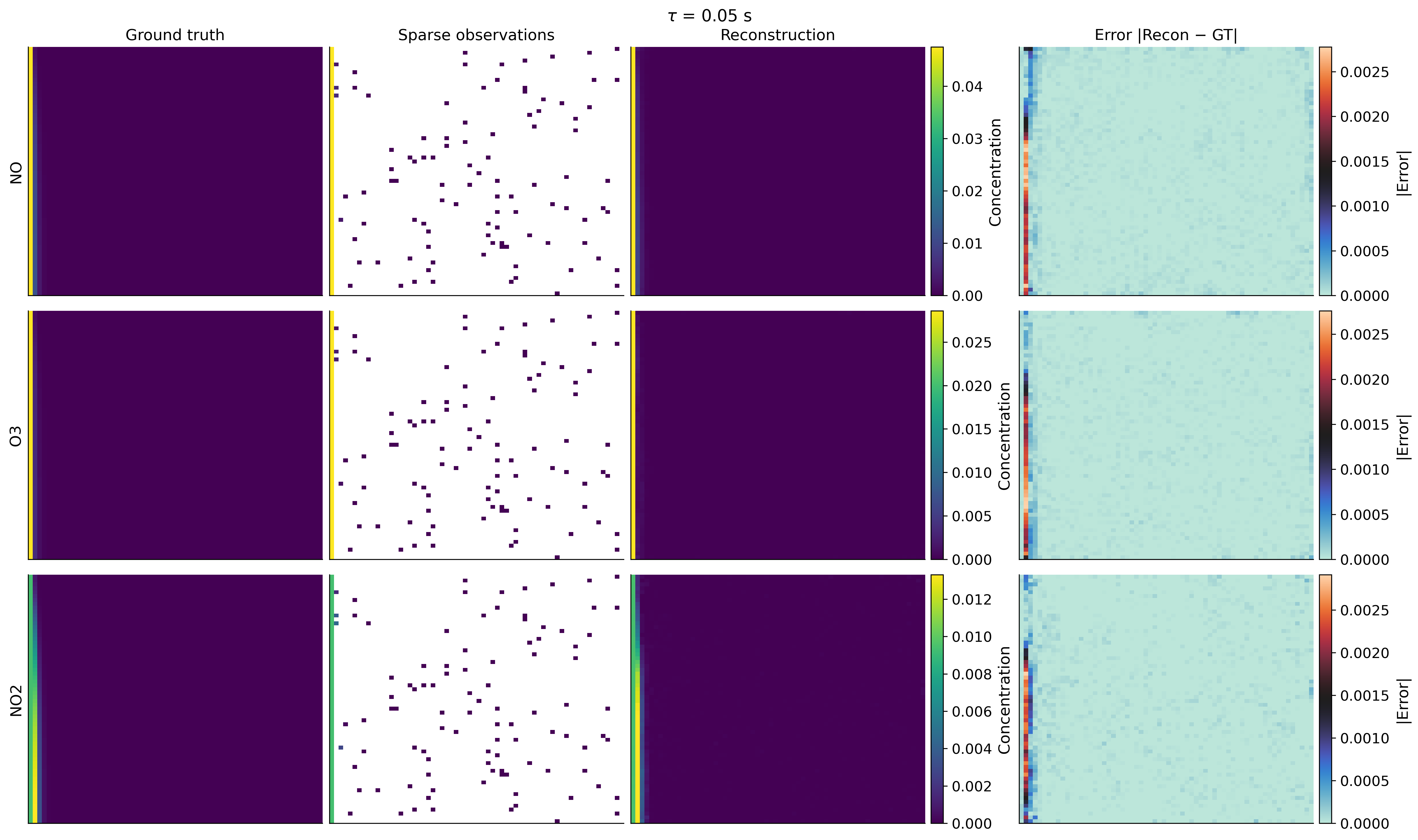}
  \caption{$\tau=0.05$}
\end{subfigure}

\vspace{0.6em}

\begin{subfigure}{\textwidth}
  \centering
  \includegraphics[width=0.8\linewidth]{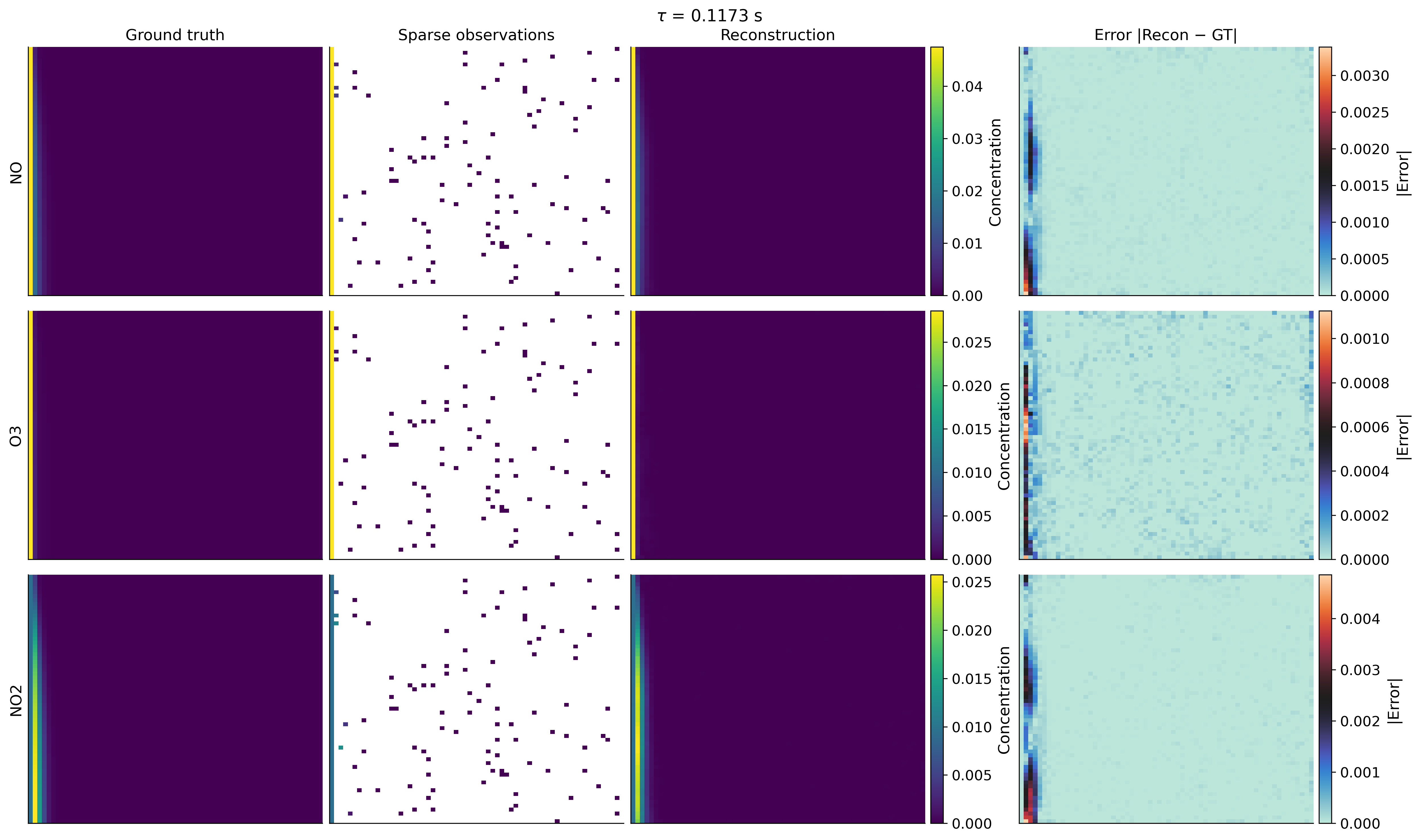}
  \caption{$\tau=0.1173$}
\end{subfigure}

\caption{Reconstruction fields and errors at different time points.}
\label{fig:triptych_time_evolution}
\end{figure}

\begin{figure}[t]\ContinuedFloat
\centering

\begin{subfigure}{\textwidth}
  \centering
  \includegraphics[width=0.8\linewidth]{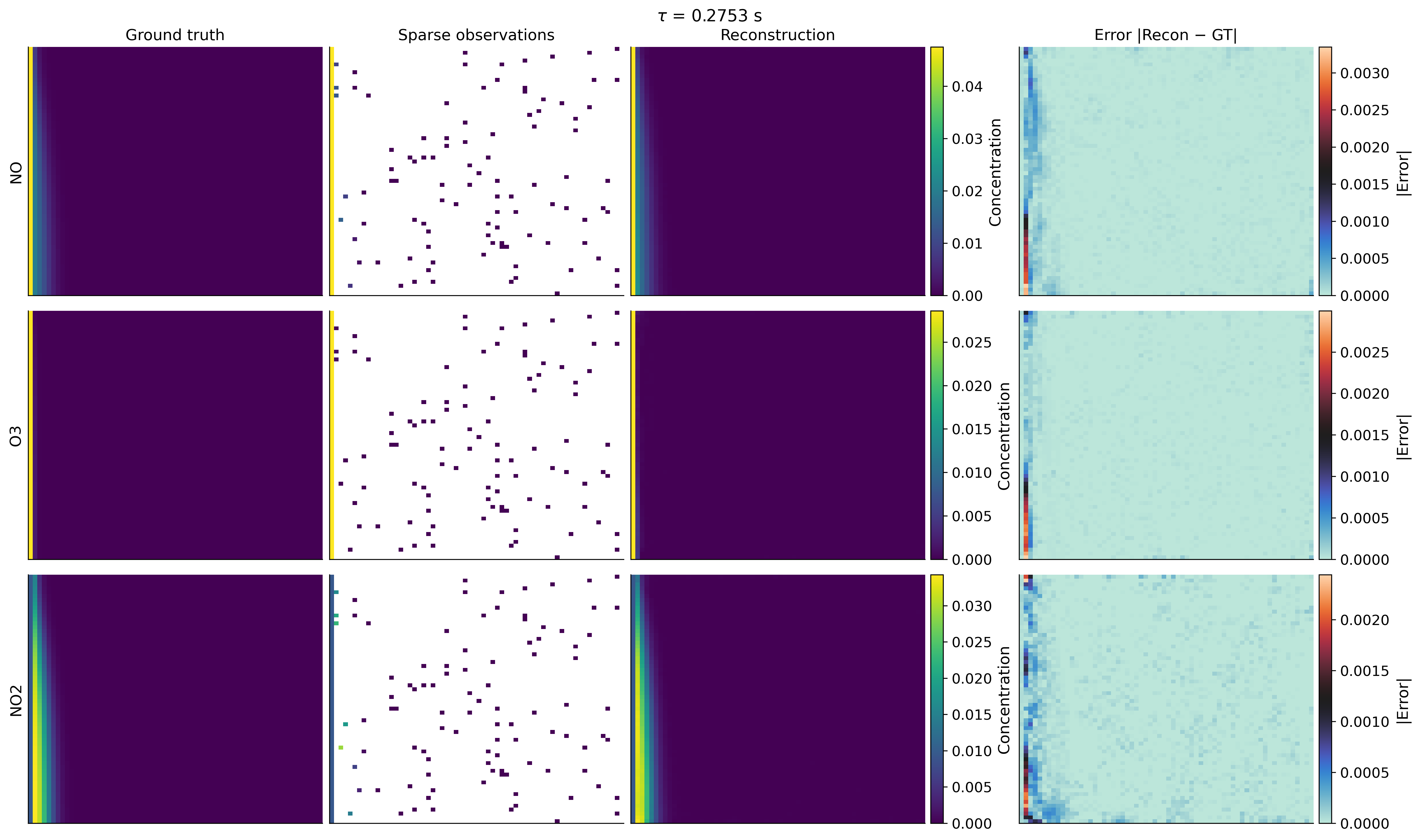}
  \caption{$\tau=0.2753$}
\end{subfigure}

\vspace{0.6em}

\begin{subfigure}{\textwidth}
  \centering
  \includegraphics[width=0.8\linewidth]{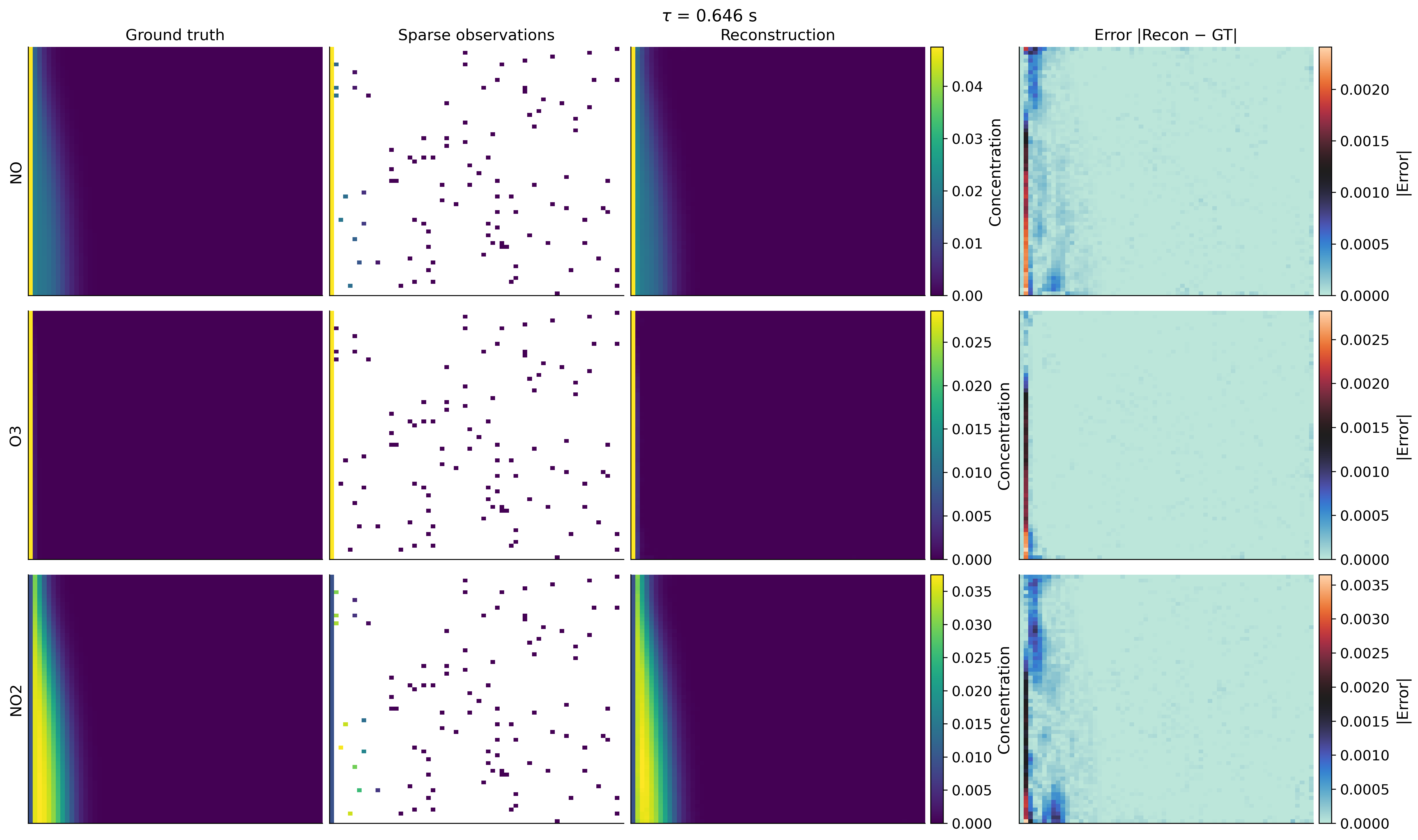}
  \caption{$\tau=0.646$}
\end{subfigure}

\vspace{0.6em}

\begin{subfigure}{\textwidth}
  \centering
  \includegraphics[width=0.8\linewidth]{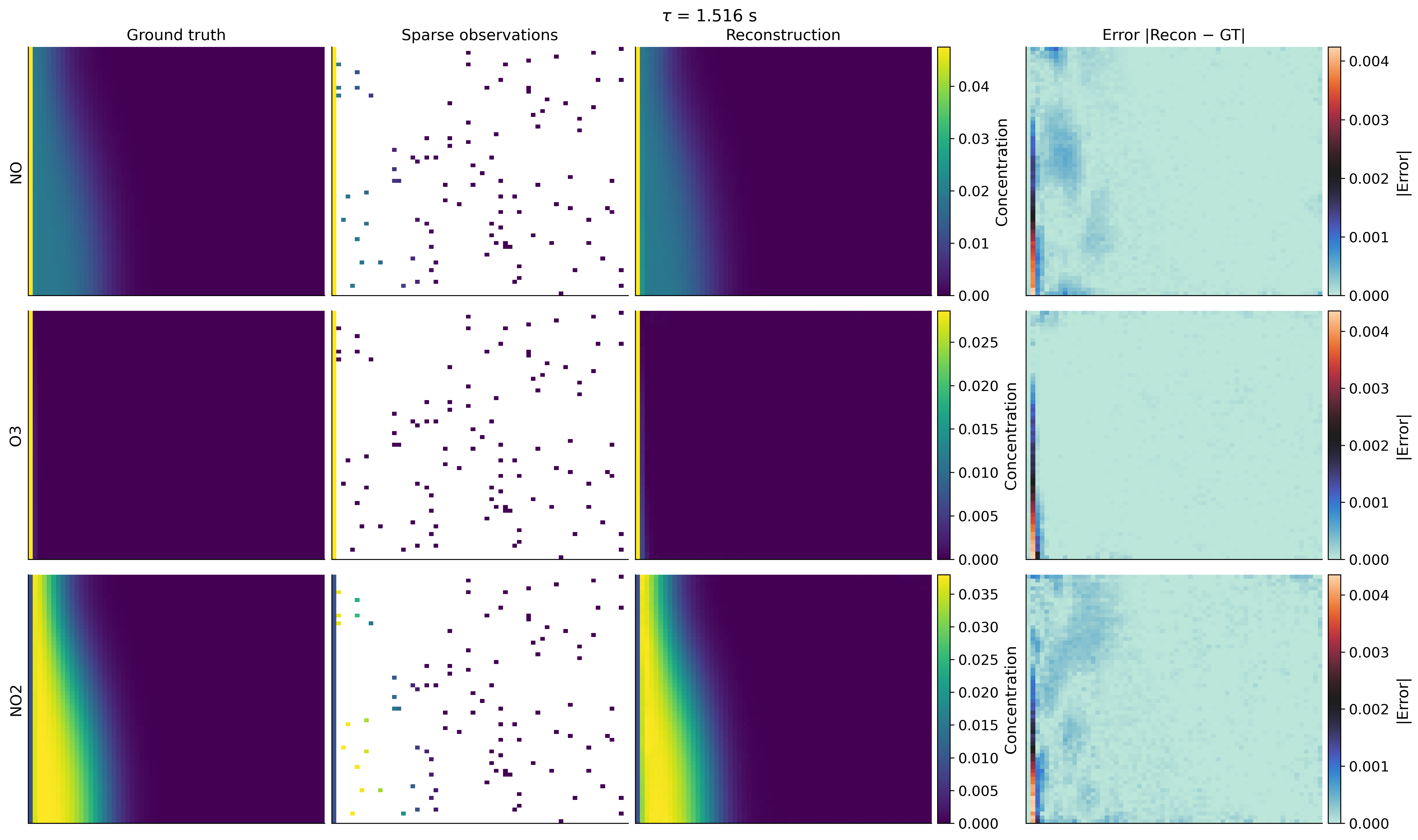}
  \caption{$\tau=1.516$}
\end{subfigure}

\caption{Reconstruction fields and errors at different time points (continued).}
\end{figure}

\begin{figure}[t]\ContinuedFloat
\centering

\begin{subfigure}{\textwidth}
  \centering
  \includegraphics[width=0.8\linewidth]{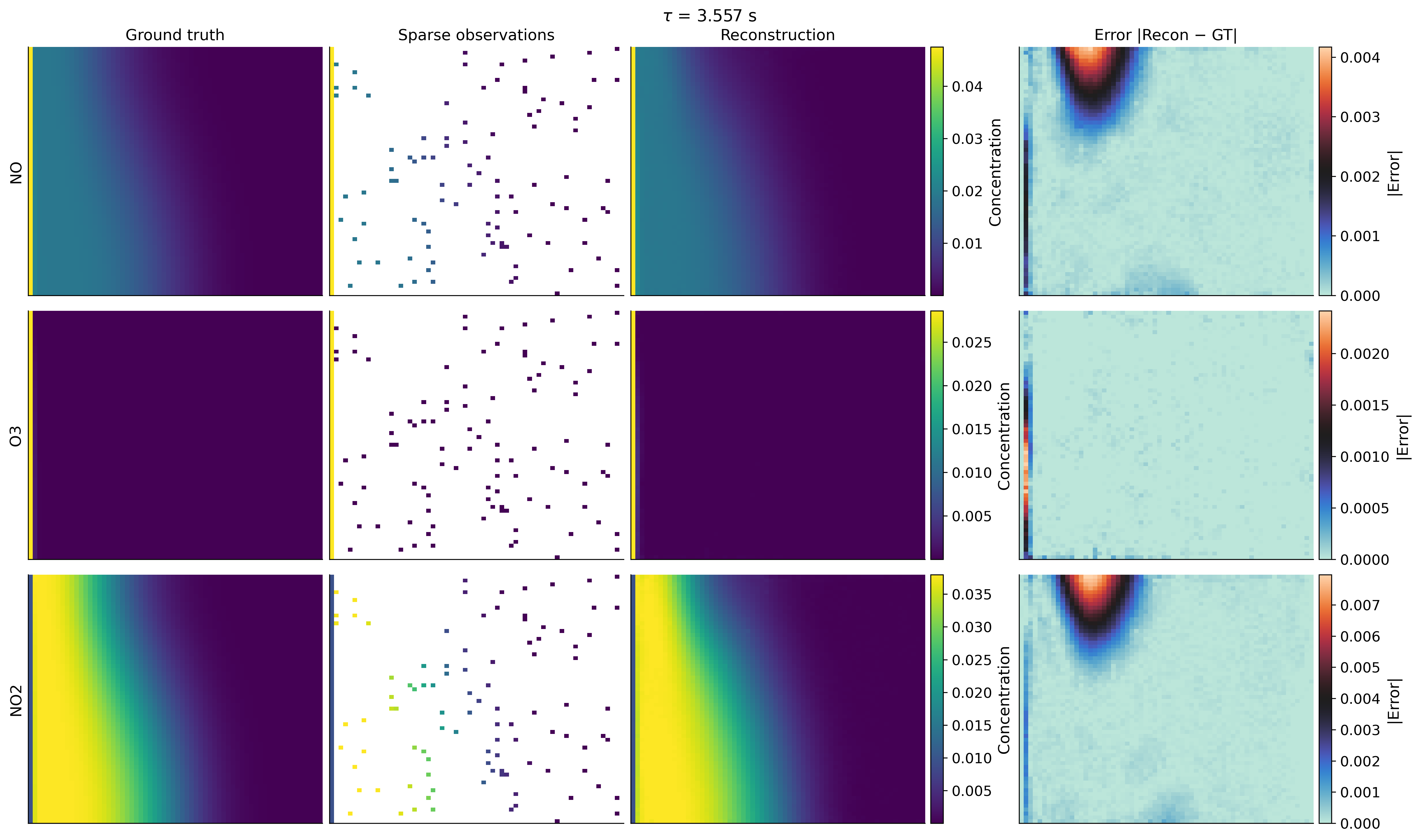}
  \caption{$\tau=3.557$}
\end{subfigure}

\vspace{0.6em}

\begin{subfigure}{\textwidth}
  \centering
  \includegraphics[width=0.8\linewidth]{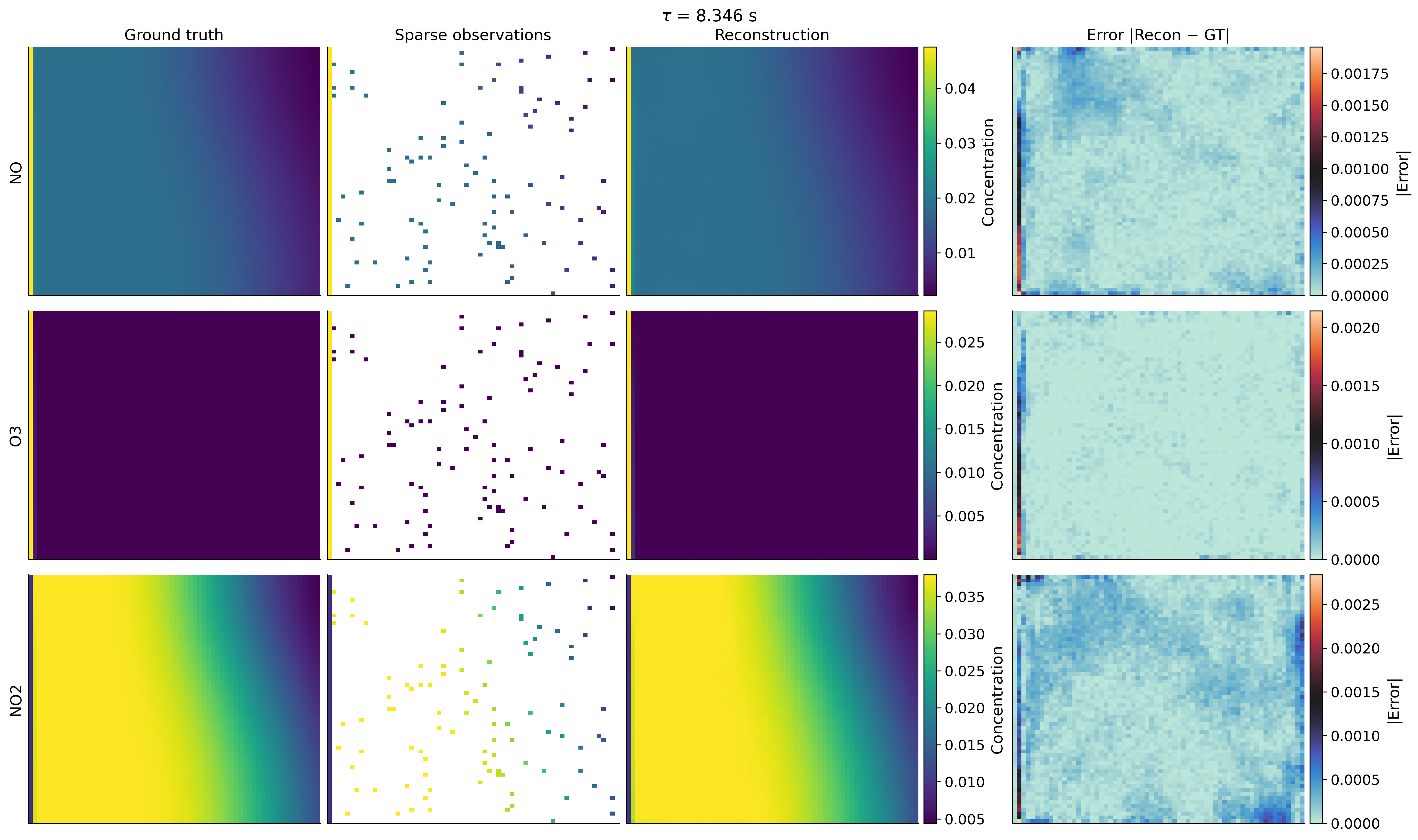}
  \caption{$\tau=8.346$}
\end{subfigure}

\vspace{0.6em}

\begin{subfigure}{\textwidth}
  \centering
  \includegraphics[width=0.8\linewidth]{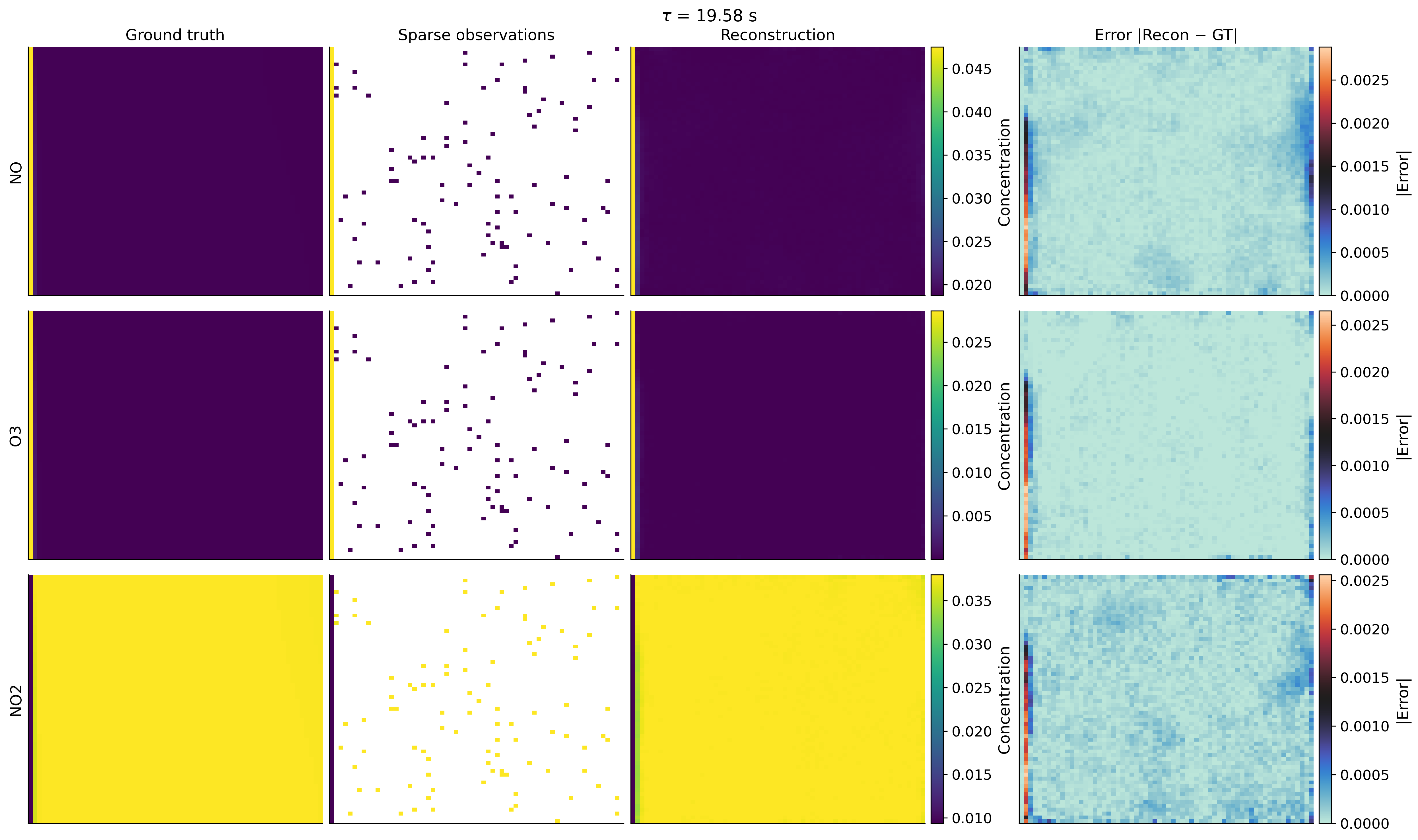}
  \caption{$\tau=19.58$}
\end{subfigure}

\caption{Reconstruction fields and errors at different time points (continued).}
\end{figure}

\end{document}